\documentclass[intlimits,twoside,a4paper]{article}

\usepackage[cp1251]{inputenc}

\usepackage[eqsecnum]{cmpj3}

\usepackage{bm}

\usepackage{mathabx}												 
\usepackage{fancyhdr}	
\usepackage{physics}

\issue{2018}{21}{3}{33301}
\doinumber{10.5488/CMP.21.33301}

\title[Two-orbital Hubbard model vs spin $S=1$ Heisenberg model]%
{Two-orbital Hubbard model vs spin $S=1$ Heisenberg model: studies on clusters}
\author{R. Lema\'{n}ski, J. Matysiak}
\address{
Institute of Low Temperature and Structure Research, Polish Academy of Science,\\ ul. Ok\'{o}lna 2, 50-422 Wroc\l aw, Poland
}

\date{Received June 15, 2018, in final form August 13, 2018}

\begin{document}

\maketitle

\begin{abstract}
We perform exact numeric calculations for the two-orbital Hubbard model on the four-site cluster. 
In the limit of large on-site coupling the model becomes equivalent to the spin $S=1$ Heisenberg model.
By comparing energy spectra of these two models, we quantified the range of interaction parameters for which the Heisenberg model satisfactorily reproduces the two-orbital Hubbard model.
Then we examined how the spectrum evolves when we are outside of this region, focusing especially on checking of how it is modified when various ways of interatomic hoppings of electrons between different orbitals are taken into account. We finally show how these modifications affect the dependence of specific heat on temperature.

\pacs 31.15.vq, 75.10.Jm, 75.10.Pq
\end{abstract}

\section{Introduction}
Multi-orbital Hubbard models (MOHMs) are suitable for studying correlated electron materials with orbital degeneracies, such as  transition metal compounds \cite{Chao3,Castellani,Pizarro}. In particular, they seem to be relevant in the theoretical analysis of single molecular magnets (SMMs). 
However, magnetic molecules have been usually described by phenomenological spin model Hamiltonians with the dominant Heisenberg interaction term (e.g., \cite{Affronte,Kozlowski}). 
This is because the low-energy excited states observed in these systems are often similar to the energy spectrum of the Heisenberg model.
Then, attempts were made to describe these systems from the first principles using the DFT method
and to determine the exchange parameters entering the Heisenberg model from the comparison of energies of various magnetic configurations \cite{Brzostowski}. The resulting energy spectrum
obtained in these calculations slightly deviates from the spectrum of the simplest version of the Heisenberg
model, but it can be well tuned to the spectrum of the Falicov-Kimball model \cite{Brzostowski}.
Unfortunately, the calculations reported in \cite{Brzostowski} did not take into account the spin flip processes, which made the exchange integrals  too large to be compatible with the experimental data.

It has been recently proposed to describe SMMs using the MOHM in the large interaction limit combined with DFT calculations \cite{Chiesa}. In this work, the values of exchange integrals were obtained within the perturbation theory. They turned out to be several times smaller than those obtained in \cite{Brzostowski}, thus better suited to experimental data.
The point is that in \cite{Chiesa}, the DFT was not used to calculate the energies of the system with different magnetic configurations, but it was focused on determining the local Coulomb and Hund interactions and the amplitudes of inter-ionic hoppings of electrons. Then, the exchange integral was determined by putting these data into the formula obtained from the perturbation theory in the limit of large values of the local Coulomb interaction, when the MOHM reduces to the Heisenberg model. In fact, the transition from the Hubbard model to the spin model by applying the perturbation theory is a commonly used procedure in correlated electron systems not only in relation to magnetism, but also relative to other phenomena, such as superconductivity  \cite{Stasyuk,MacDonald,Clevland,Chao1,Chao2,Micnas,Manouskais}.

Although the approach used in \cite{Chiesa} improved the procedure of finding the integrals of the exchange,  still the system was described only using the Heisenberg model. Therefore, nothing was known what exactly we could gain when instead of the Heisenberg model we examine the system using the MOHM. Indeed, the procedure applied there did not allow one to determine the range of the model parameters, beyond which the energy spectra of the MOHM and the Heisenberg model clearly differ from each other. Moreover, regardless of the orbitals between which the electrons jump from site to site, the perturbation theory always leads to the Heisenberg model, modifying only the value of the exchange constant. In other words, the perturbative theory does not distinguish whether the electron jumps between the same or between different orbitals (`hybridization').

To find out what is the advantage of the description of a SMM using the microscope model MOHM over the phenomenological description using the Heisenberg model, and in particular to see how the energy spectra of these two models differ from each other,
here we analyze the four-site ring with two orbitals per site. Such a small ring was chosen to make it possible to perform exact numerical calculations efficiently and rather easily. Therefore, our analysis does not focus on a particular material, but rather on capturing  a role played by various hopping amplitudes of electrons, when the system goes out of the range where the second order perturbation theory is justified. For a better illustration of our results, in this work we also present a set of curves of the specific heat versus temperature for various relationships between amplitudes of electron hoppings between different orbitals.

\begin{figure}[!b]
\centering
\includegraphics[width=0.5\textwidth]{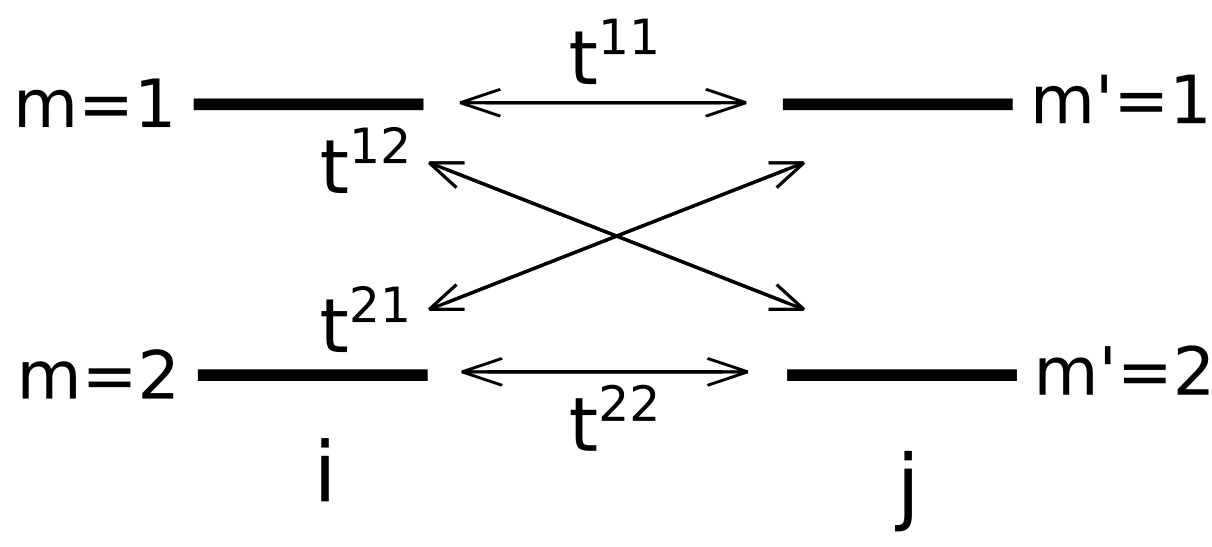}
\caption{Labelling of the hopping amplitudes $t_{mm'}$ between the sites $i$ and $j$ in the TOHM. 
$m,m'$ indicate orbital indices. }
\label{fig1}
\end{figure}

\section{Two-orbital Hubbard model (TOHM)}
Studies of the two-orbital case require an inclusion into the single-orbital Hubbard model of additional on-site couplings between various orbitals: the direct and exchange Coulomb interactions that ensure the fulfillment of the Hund's rule.
An additional complication that arises here is the possibility of electron jumping between more than one orbital and also between different orbitals (mixing terms or `hybridization') belonging to the neighboring sites (see figure \ref{fig1}).
Taking into account all these factors ensures not only an increase of the number of model parameters, but above all a significant increase of the number of states of the system, which obviously complicates diagonalization of the Hamiltonian.

The Hamiltonian $H_{\text{HM}}$ of the TOHM that we study in this paper is as follows:
\begin{equation}
\begin{aligned}
H_{\text{HM}}&=H_{0}+H_{1}\,,\\
H_{0}&=U\sum\limits_{im} n_{im \downarrow}n_{im \uparrow}
+ \frac{1}{2} \sum\limits_{i,m \neq m',\sigma} \left( U'n_{im\sigma}n_{im'\bar{\sigma}}+U''n_{im\sigma}n_{im'\sigma} \right)\\
&\quad+ \frac{1}{2} \sum\limits_{i,m \neq m',\sigma} \left( 
J c_{im\sigma}^\dagger c_{im'\bar{\sigma}}^\dagger c_{im\bar{\sigma}} c_{im'\sigma}+  
J c_{im\sigma}^\dagger c_{im\bar{\sigma}}^\dagger c_{im'\bar{\sigma}} c_{im'\sigma}
\right), \\
H_{1}&=\sum\limits_{i \neq j,m,m',\sigma}t_{mm'}c_{im\sigma}^\dagger c_{jm'\sigma}\,, 
\end{aligned}
\label{HamHubbard}
\end{equation}
where $i$ and $j$ denote nearest-neighbour sites, $m,m'$ label orbitals and $\sigma,\bar{\sigma}$ label spins of electrons ($\bar{\sigma}=-\sigma$). $U$, $U'$ and $U''$ describe the Coulomb type on-site interactions between two electrons: $U$~--- on the same orbital and $U'$ ($U''$) --- on different orbitals with opposite (parallel) spins, respectively. $J$~represents the on-site exchange coupling, but it also enters the interaction constants in the TOHM. Indeed, we take $U'=U-2J$, $U''=U-3J$, which is valid for $T_{2g}$ orbitals in an octahedral crystal field~\cite{Fresard}.

\begin{figure}[!t]
\centering
\includegraphics[width=0.55\textwidth]{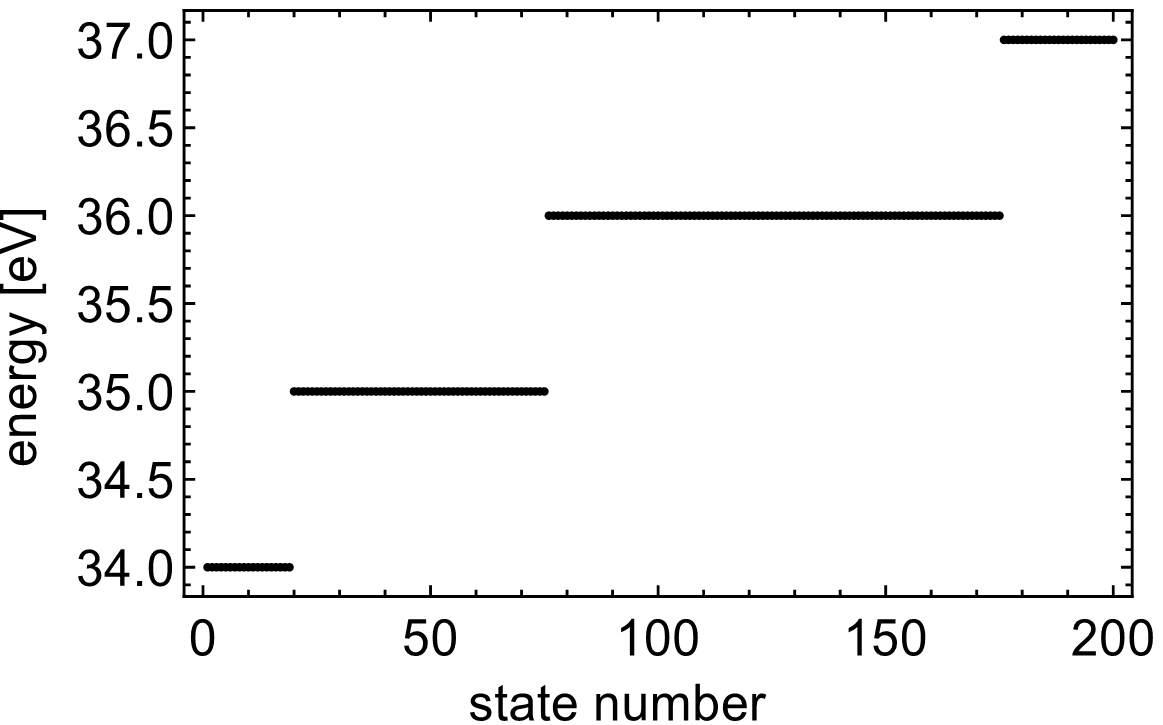}
\caption{A part of energy spectrum of the half-filled TOHM for the four-site ring in the atomic limit. The interaction parameters are $U=10$~eV, $J=0.5$~eV, $S_z=0$.}
\label{totalSpectrum}
\end{figure}

The Hamiltonian (\ref{HamHubbard}) preserves the total magnetisation. Therefore, in the beginning we consider only the states with total magnetisation equal to zero, which constitutes the largest subspace. The total number of these states is 4900. It is helpful to analyse first the spectrum of the TOHM in the atomic limit, i.e., when all $t_{mm '}=0$ (the lowest part of the spectrum is shown in figure \ref{totalSpectrum}). We have chosen the parameters in our calculations to be $U=10$~eV, $J=0.5$~eV. The spectrum consists of 26 degenerate energy levels. Then, we are interested in the region of small values of the hopping constants $t_{mm'}$, and it is understood that the overall structure of the spectrum in figure \ref{totalSpectrum} is preserved, but each of the degenerate levels splits in a specific way when $t_{mm'}$ becomes non-zero. Here, we focus mainly on finding the splitting of the lowest energy level.

To obtain the exact spectrum of the model~(\ref{HamHubbard}) for our small system, we use the exact diagonalization method.

\section{Computation details}
In the atomic limit,  the  eigenstates of the Hamiltonian~(\ref{HamHubbard}) for a single site together with their representations and energies are shown in table~\ref{tableSingleSite}. As indicated in the first column, they are also the eigenstates  of the total spin $S$ and  its $z$ component $S_z$. The full Hilbert space base for the whole 4-site system is formed from the tensor products of the states from table~\ref{tableSingleSite}.  

For a site occupied by two electrons, the lowest energy refers to the eigenstates 1--3 from table~\ref{tableSingleSite}. However, obviously,  after one of the electrons jumps from one site to another, the first site will contain only one electron and the other one three electrons. Then, the lowest energy eigenstate for the site with one electron  is one of the eigenstates 7--10 from table~\ref{tableSingleSite}., and for the site with three electrons, it is one of the eigenstates 11--14 from table~\ref{tableSingleSite}.  

\begin{table}[!t]
\caption{Eigenstates and eigenvalues of the model (\ref{HamHubbard}) for a single site with two orbitals per site. 
Sections separated by double horizontal lines correspond to the different numbers of electrons $n$ on the site.}
\label{singleEigenstates}
\vspace{2ex}
\centering
\begin{tabular}{| c | l | c | c |}
\hline
\hline
nr&$\ket{n,S,S_z}$& State representation & Energy \\ \hline
\hline
1&$\ket{2,1,1}$ &  $\qty(\uparrow | \uparrow)$ & $U-3J$ \\ \hline
2&$\ket{2,1,0}$ &  $\dfrac{1}{\sqrt{2}} \qty\Big( \qty(\uparrow | \downarrow) + (\downarrow | \uparrow))$ & $U-3J$ \\ \hline
3&$\ket{2,1,-1}$ &  $\qty(\downarrow | \downarrow)$ & $U-3J$ \\ \hline
4&$\ket{2,0,0}$ &  $\dfrac{1}{\sqrt{2}} \qty\Big( \qty(\uparrow | \downarrow) - (\downarrow | \uparrow))$ & $U-J$ \\ \hline
5&$\ket{2,0_a,0}$ &  $\dfrac{1}{\sqrt{2}} \qty\Big(  \qty(\updownarrows | -)+( - |\updownarrows ))$ & $U-J$ \\ \hline
6&$\ket{2,0_b,0}$ &  $\dfrac{1}{\sqrt{2}} \qty\Big(  \qty(\updownarrows | -)-( - |\updownarrows ))$ & $U+J$ \\ \hline
\hline
7&$\ket{1,1/2_a,1/2}$ &  $\qty(\uparrow | - )$ & $0$ \\ \hline
8&$\ket{1,1/2_a,-1/2}$ &  $\qty( \downarrow | -)$ & $0$ \\ \hline
9&$\ket{1,1/2_b,1/2}$ &  $\qty( - | \uparrow )$ & $0$ \\ \hline
10&$\ket{1,1/2_b,-1/2}$ &  $\qty( - | \downarrow)$ & $0$ \\ \hline
\hline
11&$\ket{3,1/2_a,1/2}$ &  $\qty(\uparrow | \updownarrows  )$ & $3U-5J$ \\ \hline
12&$\ket{3,1/2_a,-1/2}$ &  $\qty( \downarrow | \updownarrows )$ & $3U-5J$ \\ \hline
13&$\ket{3,1/2_b,1/2}$ &  $\qty( \updownarrows  | \uparrow )$ & $3U-5J$ \\ \hline
14&$\ket{3,1/2_b,-1/2}$ &  $\qty( \updownarrows  | \downarrow)$ & $3U-5J$ \\ \hline
\hline
15&$\ket{0,0,0}$ &  $ \qty( - | - )$ & $0$ \\ \hline
\hline
16&$\ket{4,0,0}$ &  $\qty( \updownarrows  |\updownarrows )$ & $6U-10J$ \\ \hline
\hline
\end{tabular}
\label{tableSingleSite}
\end{table}

In the case of a ring with 4 sites and 2 orbitals per site, we have the total number of orbitals $r= 8$. Then, at half filling, i.e.,   when the number of electrons is equal to the namber of orbitals, the numbers of the states that are needed to diagonalize the Hamiltonian are: 12870 for the TOHM, but only  $3^4=81$ for the Heisenberg model.

\subsection{Perturbation theory}

In the  perturbation theory, $H_{1}$ is taken as the perturbation and $H_{0}$ is the Hamiltonian for which the solution is known. Contrary to the single orbital model, in the multi-orbital case, the states with one electron per orbital and with different spin configurations have, in general, different energies even in the limit when $t_{mm'}=0$ for all hopping amplitudes. Then, in order to obtain an effective Hamiltonian with only spin degrees of freedom, we  restrict our considerations to a single subspace, where all unperturbed states are degenerate. In particular, we focus on the lowest energy subspace.

Here, we only provide the final result without going through the calculations. A detailed derivation of the formulae reported below can be found in \cite{Chiesa}. The result, which is valid only for the subspace with the lowest energy, is that the effective Hamiltonian turns out to be the standard Heisenberg Hamiltonian with the effective Heisenberg constant $\Gamma_{\text{eff}}$:

\begin{equation}
\begin{aligned}
H_{\text{eff}}&=\Gamma_{\text{eff}}\sum_{i \neq j}\qty(S_iS_j-4),\\
\Gamma_{\text{eff}}&=\dfrac{4\abs{t_{\text{eff}}}^2}{U+J}.
\end{aligned}
\label{Heff}
\end{equation}

$\Gamma_{\text{eff}}$ is determined by a single effective hopping constant $t_{\text{eff}}$, which in a simple way depends on an initial hopping constants $t_{mm'}$:

\begin{equation}
t_{\text{eff}}=\dfrac{\sqrt{\sum_{m,m'}t_{mm'}^2}}{2}.
\label{teff}
\end{equation}

\section{The lowest part of energy spectra}

\begin{figure}[!t]
\centering
\includegraphics[width=0.55\textwidth]{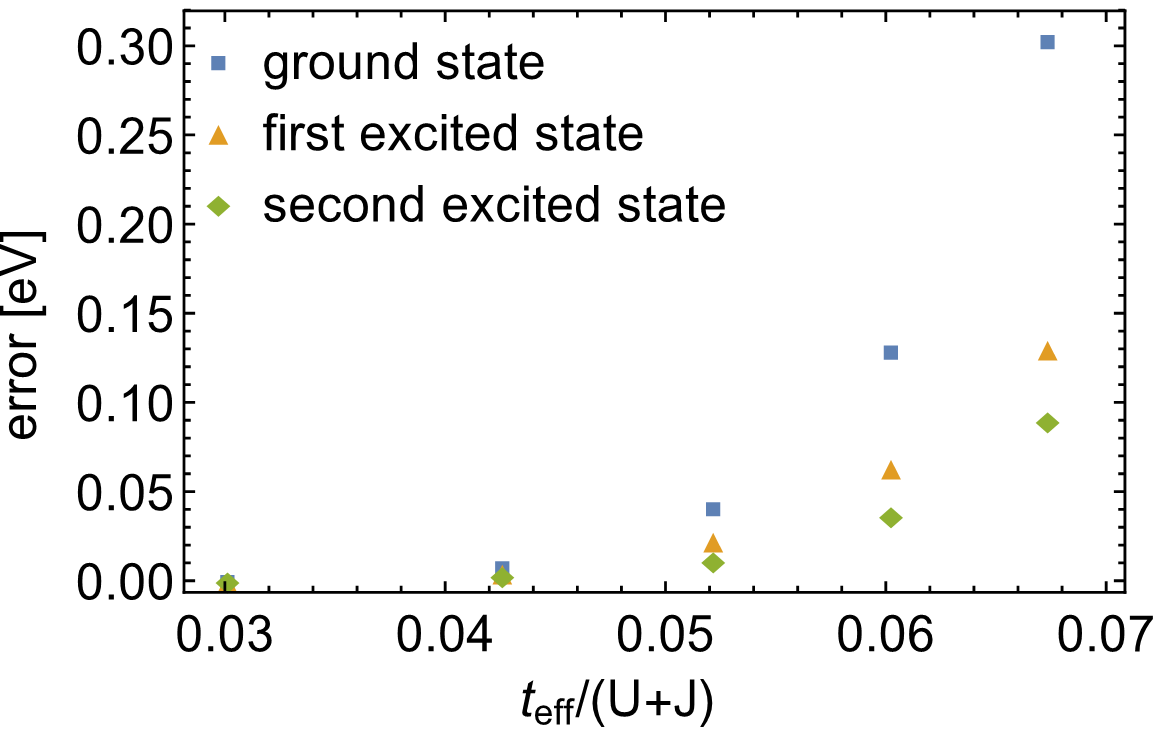}
\caption{(Colour online) Comparison of the difference between the energies calculated for  the  TOHM and for the Heisenberg model as a function of $t_{\text{eff}}/(U+J)$ ($S_z=0$, $t_{12}=t_{21}=0$, $t_{11}=t_{22}$). The data are shown for the lowest three eigenvalues. The interaction parameters are: $U=10$~eV, $J=0.5$~eV.}
\label{methods_comparision}
\end{figure}

We start the presentation of our results by specifying the conditions for the model parameters for which the perturbation theory is justified. Our calculations are illustrated in figure \ref{methods_comparision} where the differences between the energies of the ground states, of the first excited states and of the second excited states were determined from the exact diagonalization and by using the Heisenberg model with the effective exchange integral calculated from the perturbation theory. 
From figure \ref{methods_comparision} it is clear that these differences increase more steeply than linearly with an increase of the ratio $t_{\text{eff}}/(U + J)$. Interestingly, the energies of the ground state differ the most, almost twice less differ the energies of the first excited state and even less do the ones of the second excited state.

Obviously, the limit value of $t_{\text{eff}}/(U + J)$ above which the Heisenberg model ceases to describe this system adequately, can be fixed arbitrarily. Here, we assumed that the difference between the state energies should not exceed 0.03 eV. This corresponds to $t_{\text{eff}}/(U + J)$ value equal to 0.05.

\begin{figure}[!b]
\centering
\includegraphics[width=0.7\textwidth]{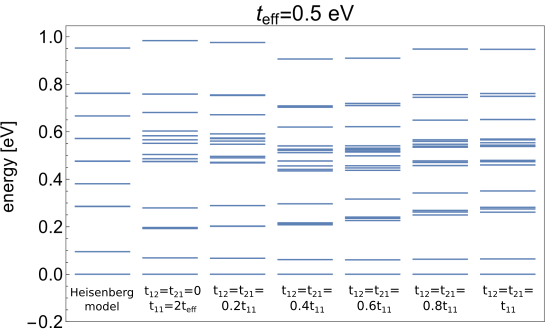}
\caption{(Colour online) Comparison of the energy spectrum of the Heisenberg model with the lowest parts of the energy spectra of the TOHM for the set  of the following relationships between the hopping amplitudes: $t_{12}/t_{11}=0, 0.2, 0.4, 0.6, 0.8, 1$.
Here, $t_{21}=t_{12}$, $t_{22}=0$, $U=10$ eV and $J=0.5$ eV, $S_z=0$. In all these cases, the same $t_{\text{eff}}=0.5$ is kept. For the Heisenberg model, the exchange constant $\Gamma_{\text{eff}}=4t_{\text{eff}}^2/(U+J)$.}
\label{d2}
\end{figure}

Next, we present the evolution of the lowest part of the energy spectrum of the TOHM depending on the ratio $t_{12}/t_{11}$ between the  hopping amplitudes of electrons, while the constant value of $t_{\text{eff}}$ is maintained. In the case presented in figure \ref{d2}, we have chosen $t_{\text{eff}}=0.5$.
When $t_{\text{eff}}$ is constant, then the perturbation theory gives the energy spectrum typical of the Heisenberg model, which does not depend on the ratio $t_{12}/t_{11}$ (see the first column in figure \ref{d2}). However, the exact solution  shows that the energy spectrum of the TOHM clearly depends on the ratio $t_{12}/t_{11}$, which is illustrated by the other columns in figure \ref{d2}. Thus, the theory of perturbation is too imprecise to distinguish the cases with different ways of electron jumping between neighboring magnetic ions.
In fact, in figure \ref{d2}, we present only the energy levels with the total $S_z=0$.

\begin{figure}[!t]
\centering
\includegraphics[width=0.55\textwidth]{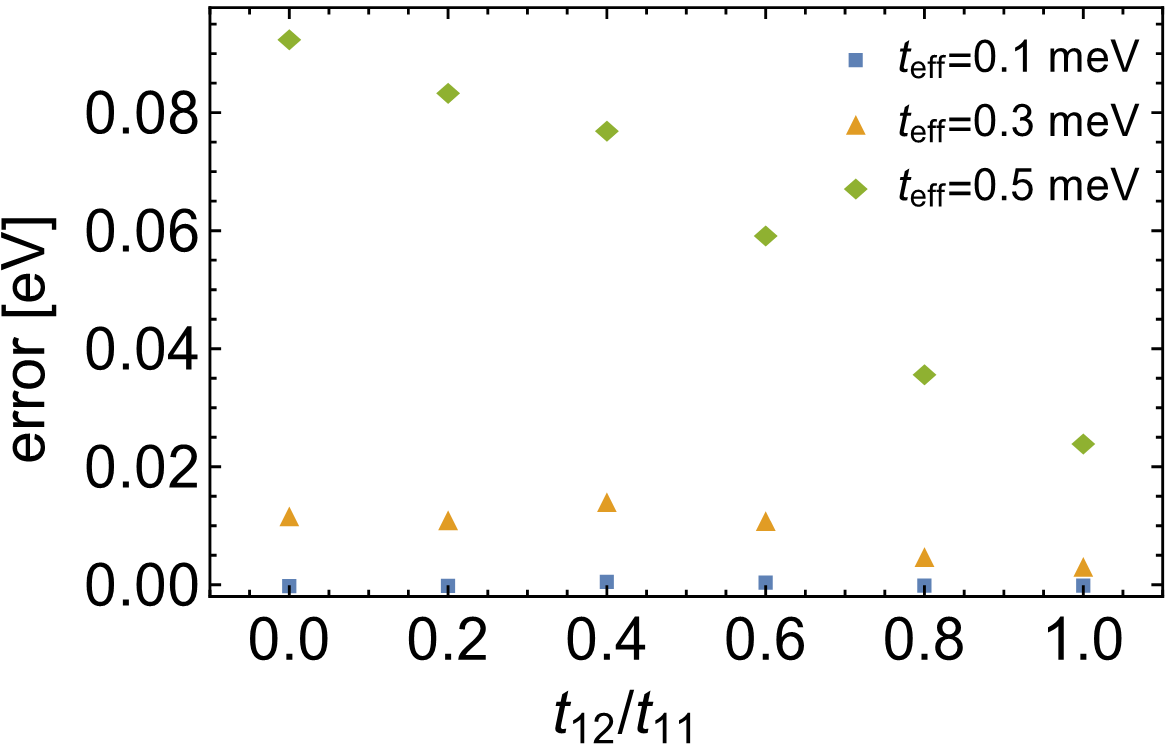}
\caption{(Colour online) Comparison of the difference between energies  of the second excited state calculated exactly for  the  MOHM and for the Heisenberg model as a function of $t_{12}/t_{11}$, $S_z=0$.  The interaction parameters are: $U=10$~eV, $J=0.5$~eV.}
\label{a5}
\end{figure}

The difference between the exact solution and that based on the Heisenberg model is clearly seen by comparing the energy of the second excited state  depending on the ratio $t_{12}/t_{11}$ for several different values of $t_\text{{eff}}$, as it is shown in figure \ref{a5}. For that particular state, the difference decreases with an increasing $t_{12}/t_{11}$ ratio. The same trend takes place for all $t_\text{{eff}}$ values, though the magnitude of the difference increases exponentially.

\begin{figure}[!b]
\vspace{-3mm}
\centering
\includegraphics[width=0.7\textwidth]{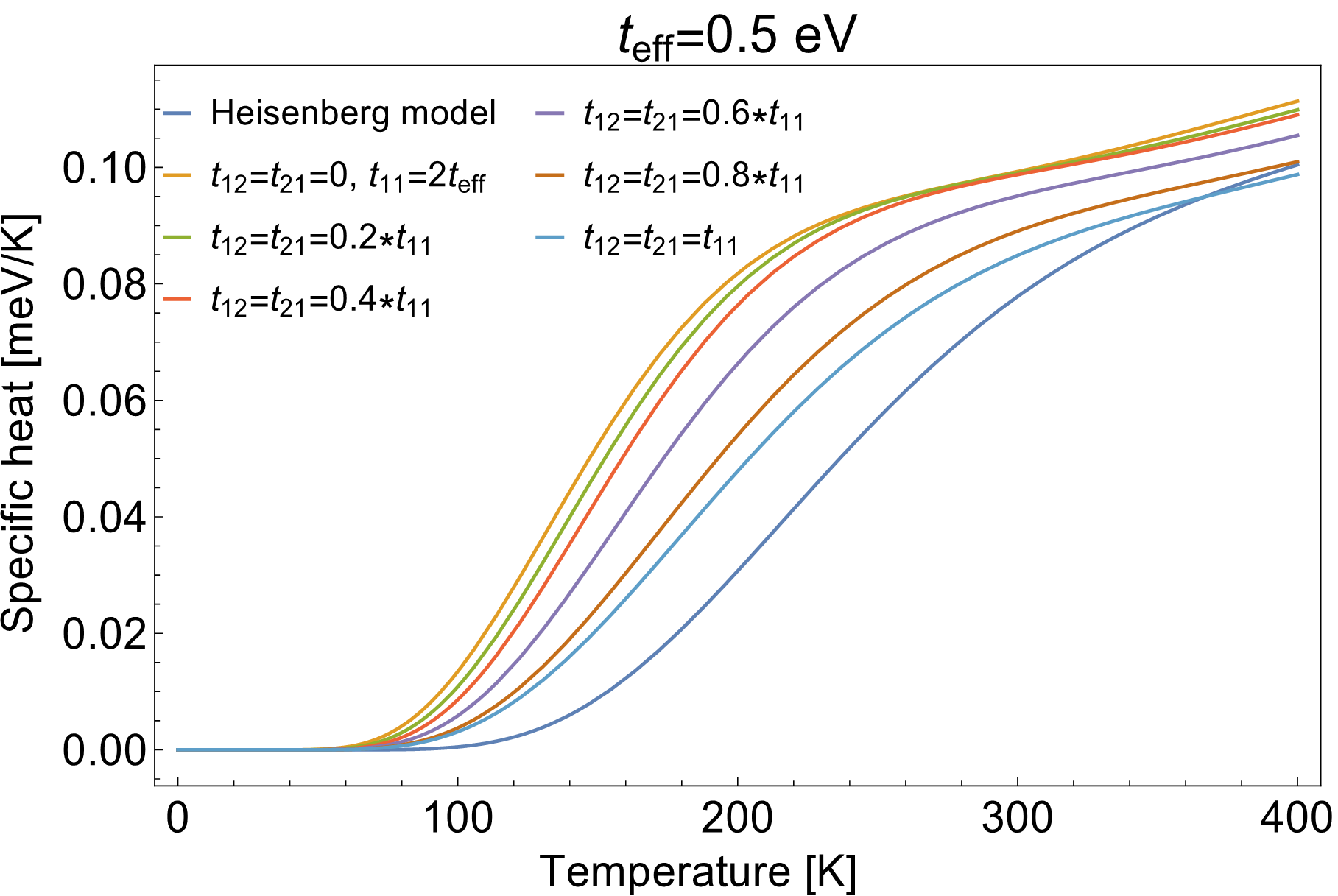}
\caption{(Colour online) Specific heat as a function of temperature for the Heisenberg model with $\Gamma_{\text{eff}}=4t_{\text{eff}}/(U+J)=0.095$~eV and for various versions of the MOHM model (for all of them  $t_{\text{eff}}=0.5$ eV is kept). The curves for the MOHM model differ by ratios of the hopping constants $t_{12}/t_{11}$ ($t_{21}=t_{12}$, $t_{22}=0$), as indicated on the plot legend. The parameters of the model are: $U=10$~eV, $J=0.5$~eV.}
\label{Cv05}
\end{figure}

\begin{figure}[!t]
\centering
\includegraphics[width=0.7\textwidth]{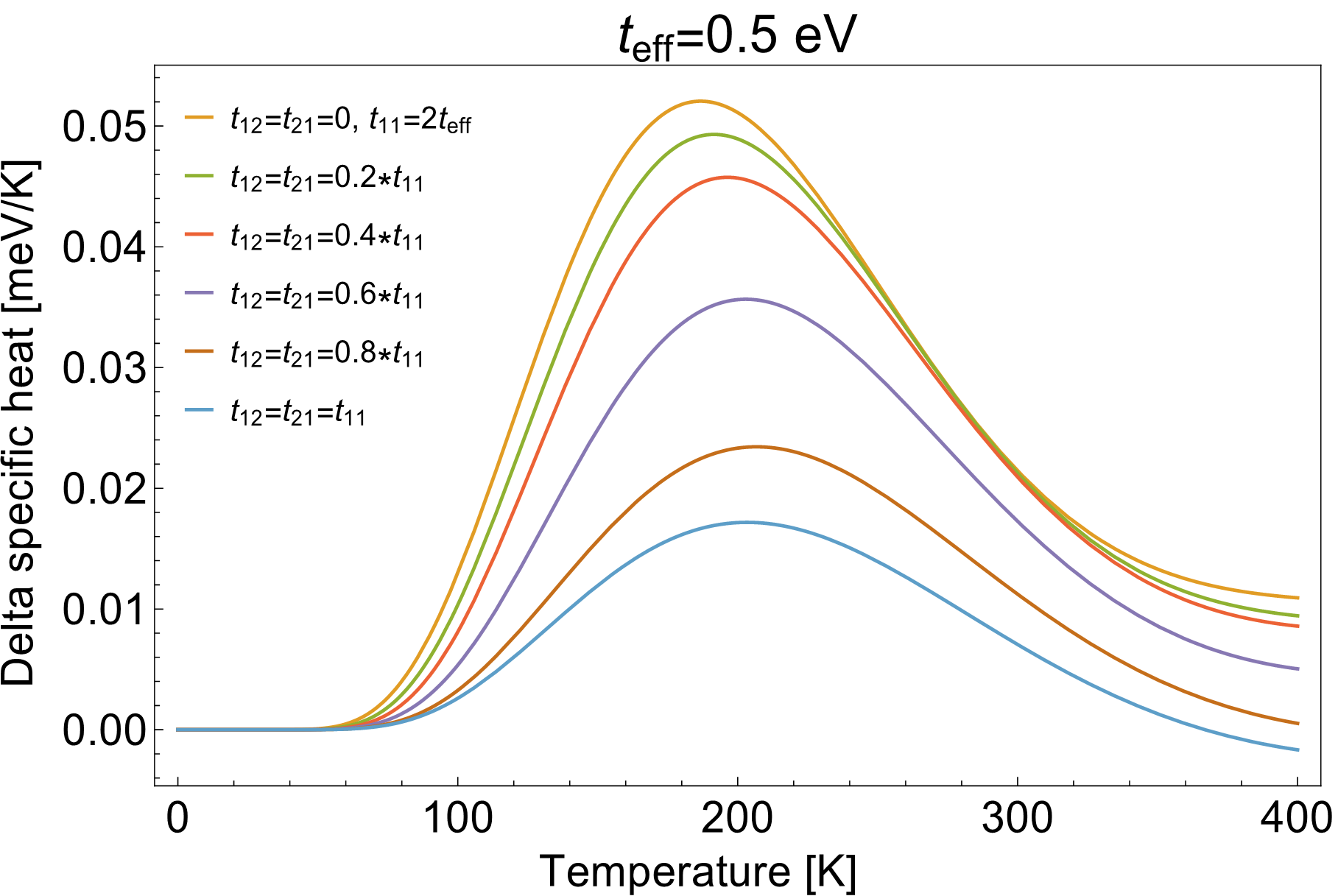}
\caption{(Colour online) The difference between specific heat of the Heisenberg model with $\Gamma=4t_{\text{eff}}/(U+J)=0.095$~eV and the different cases of the MOHM model  (for all of them  $t_{\text{eff}}=0.5$ eV is kept). The curves for the MOHM model differ by ratios of the hopping constants $t_{12}/t_{11}$ ($t_{21}=t_{12}$, $t_{22}=0$), as indicated on the plot legend. The parameters of the model are: $U=10$~eV, $J=0.5$~eV.}
\label{Cv05roznica}
\end{figure}

It is obvious that the distribution of the lowest energy levels determines the dependence of specific heat on temperature
as it is illustrated in figures \ref{Cv05} and \ref{Cv05roznica}. Figure \ref{Cv05} illustrates the $C_{v}(T)$ changes with an increase of the $t_{12}/t_{11}$ ratio for the fixed value $t_{\text{eff}}=0.5$ for the TOHM, as well as $C_v(T)$ for the Heisenberg model for comparison. In turn, in figure \ref{Cv05roznica}, the differences between the values of $C_{v}(T)$ calculated for the TOHM and the Heisenberg model with an increase of the ratio $t_{12}/t_{11}$ for the same value of $t_{\text{eff}}=0.5 $ are displayed. Additionally, table~\ref{maxima} contains numeric values of the maxima of the curves from figure \ref{Cv05roznica}. Based on the drawings given in figures \ref{Cv05} and \ref{Cv05roznica}, and especially in the table~\ref{maxima},  it can be supposed that from a course of the dependency $C_{v}$ you can draw conclusions about the degree of hybridization expressed by the ratio $t_{12}/t_{11}$.

\begin{table}[!t]
\caption{Maxima of the differences between the values of specific heat calculated for the TOHM and for the Heisenberg model calculated for various $t_{12}/t_{11}$ ratios.}
\vspace{2ex}
\centering
\begin{tabular}{|c|c|c|c|c|c|c|}
\hline\hline
$t_{12}/t_{11}$ ratio& 0 & 0.2 & 0.4 & 0.6 & 0.8 & 1  \\ \hline\hline
maximum $\Delta C_v$ [meV/K] &0.051& 0.049 & 0.046 & 0.036 & 0.024 & 0.018 \\
\hline\hline
\end{tabular}
\label{maxima}
\end{table}

\section{Summary and conclusions}
We carried out exact numerical calculations for the four-site ring with two orbitals and two electrons per site, using the TOHM and we compared the results with those obtained for the respective Heisenberg model. On the basis of this comparison, we estimated the range of the interaction parameters, and, more specifically, the range of the $t_{\text{eff}}/(U + J)$ ratio, for which the TOHM is well represented by the Heisenberg model.

However, the main achievement of our work is a demonstration that the energy spectrum of the system depends in a significant way on how an electron can jump from one site to another. Indeed, the case when an electron can only jump between the same type of orbital is different from the case when the electron can jump both between the same type and between different types of orbitals. This difference manifests itself through a modification of the energy spectrum, which evidently evolves with an increase in the degree of hybridization between different orbitals. This, in turn, brings a change, e.g., into the dependence of specific heat on temperature $C_{v}(T)$. It can, therefore, be supposed that precise measurements of the specific heat would enable us to determine the degree of hybridization between different orbitals located on neighboring sites. However, to make this possible, the ratio of the effective amplitude $t_{\text{eff}}$ of the electron jump to the on-site interaction constant $U+J$ should be not very small, because only in that case the TOHM is significantly different from the Heisenberg model.

\ukrainianpart 

 \title{Порівняння двоорбітальної моделі Хаббарда та спін $S=1$ моделі 
 Гайзенберга: дослідження на кластерах} 
 \author{Р. Лєманьскі, Я. Матисяк} 
 \address{ 
     Інститут низьких температур та структурних досліджень Польської академії 
 наук,\\ вул. Окульна 2, 50-422 Вроцлав, Польща 
 } 

 \makeukrtitle 

 \begin{abstract} 
 \tolerance=3000%
 Нами проведено точні числові розрахунки для двоорбітальної моделі Хаббарда на 
 чотиривузловому кластері. В границі великої одновузлової взаємодії ця модель 
 стає еквівалентною спін $S=1$ моделі Гайзенберга. З порівняння енергетичних 
 спектрів цих двох моделей нами визначено кількісно діапазон значень параметрів 
 взаємодії, для якого модель Гайзенберга задовільно відтворює двоорбітальну 
 модель Хаббарда. Потім ми розглянули зміни спектру, коли ми знаходимося за 
 межами цього діапазону, зосереджуючись, зокрема, на перевірці того, як він 
 модифікується, коли враховуються різні способи міжатомних перескоків 
 електронів між різними орбіталями. Насамкінець, нами показано як ці зміни 
 впливають на температурну залежність питомої теплоємності. 
 \keywords багатоорбітальна модель Хаббарда, модель Гайзенберга, магнітні молекули 

 \end{abstract}

\end{document}